\newif\iflayout
\layouttrue
\documentstyle[twocolumn,aps,prl,epsfig,rotating]{revtex}
\begin{document}

\protect\twocolumn [\hsize\textwidth\columnwidth\hsize\csname
@twocolumnfalse\endcsname

\title{
A  Study of the affect of N and B doping\\
on the growth of CVD diamond (100):H 2 $\times$ 1 surfaces\\}

\author{
M. Kaukonen and R. M. Nieminen
}
\address{
{\small \it Department of Physics, Technical University, FIN 02150 Helsinki, Finland}\\
}

\author{ P. K. Sitch,  G. Jungnickel, D. Porezag and Th. Frauenheim}
\address{
{\small \it Department of Physics, Technical University, D 09107 Chemnitz, Germany}
}

%\date{today}
\maketitle

\begin{abstract}
The doping of the CVD-diamond (100):H 2 $\times$ 1 surface with B and N 
has been studied using  
the density functional tight--binding method (DF--TB).
In agreement with recent experimental results,
B doping is found to lower the abstraction energies and
remove diffusion barriers along the diamond growth pathway 
proposed by Harris.
In contrast, the Harris mechanism is less favorable
with N doping, casting doubt on its validity in this case.
We therefore propose a novel growth pathway on  N-doped 
CVD-diamond (100):H 2 $\times$ 1 
surfaces.
This involves a dimer opening reaction
and requires less H abstraction reactions compared to the
Harris mechanism. \\

\end{abstract}

]

\section{introduction}
The understanding of diamond growth via the CVD process
has proved difficult
for both theorists and experimentalists alike. This is due to the 
 large number of
experimental parameters
contributing to the problem and an 
uncertainty about the growth species. 
Progress has been made on the latter by the work of 
D'Evelyn et al\cite{develyn}
who, using isotope labelling techniques,  
claim to have unequivocally identified the 
principal growth species to be CH$_{3}$. With this in mind, 
Harris\cite{harris} has 
proposed a complex mechanism for diamond growth, whose initial steps 
lead to 
the deposition of a CH$_{2}$ group at a bridge site above a
surface reconstruction bond. \\
Recently, the effect of B and N doping on the 
CVD growth process 
has produced a series of  intriguing results.
In the case of B, various workers 
have found that 
B improves the
crystalline quality of (100) CVD surfaces and 
enhances the p-type conductivity of the films\cite{nemanich,roth,hiraki}. 
Interest in the role of N in CVD diamond has been heightened by
experimental observations that N preferentially catalyses 
growth in the (100) direction\cite{koidl,giling,moustakas}.
To the authors' knowledge, no serious attempts have been made to explain 
these phenomena theoretically. Indeed, it is unclear whether these somewhat 
puzzling results are compatible with  the Harris mechanism or 
if in doping cases
a different  growth process is at work. In this 
paper, we  answer this question by  investigating the effect of
subsurface B and  N on the energetics of the Harris  mechanism.  
We find that the energies of the various growth steps are greatly 
altered, casting doubt on the applicability of the Harris method in these 
cases. We therefore  discuss a possible alternative to the initial 
steps of the process.\\
The paper is arranged as follows: in section II the 
theoretical tools used in this study are described, whilst section III
explains the first few steps of the Harris mechanism.
Section IV  contains theoretical  
results for the N and B doping on (100):H 2 $\times$ 1  
surfaces whilst
section V  includes a discussion of these results.
Section VI proposes a new model for the CVD diamond growth 
and a  conclusion is given in section VII.

\section{theoretical method and the model system}
The density functional tight--binding method
(DF--TB)
derives its name 
from its use of self--consistent density functional 
calculations for pseudo--atoms in order  to construct
transferable 
tight--binding (TB) potentials for a 
non--selfconsistent solution  of the Kohn--Sham equations
for the many body case.
It  differs from conventional tight--binding
techniques in that  there is  a systematic
way of deriving these potentials, independent of the atom type
involved. This is thus not a ``parametrisation''
as is usually meant when one talks about TB approaches.
For an in depth description, the reader is referred to
Ref.~\cite{dirk}.
The method  has been successfully applied to various scale carbon
 systems, ranging from small clusters to buckminster 
fullerenes and the bulk phase\cite{dirk}, the electronic and 
vibrational properties of (100) and (111)
surfaces\cite{koe,stern},  
amorphous carbon
systems of all densities\cite{uwe}, as well as boron
nitride\cite{jurg} 
and boron and nitrogen doping of diamond and amorphous
systems\cite{sitch}.
We have furthermore used the  
ab--initio cluster programs 
of Pederson and Jackson\cite{pederson} and Jones and
Briddon\cite{jones} to 
check selected results.
These programs are highly accurate but computationally very expensive, 
hence we are limited in these cases to very small clusters which can only
represent highly idealized surfaces. Nevertheless, these 
calculations  are useful insofar 
as they serve to verify the essential physics underpinning the results 
of our DF--TB work.\\
The  144 atom (100):H supercell with  the
2 $\times$ 1 reconstructed surface used in this 
investigation is shown in Fig.~\ref{fig1}.
It is made
up of  
eight reconstructed surface bonds and 
six layers of carbon atoms. The dangling bonds
on the 
lower surface are terminated with pseudo--hydrogen atoms. Unless
otherwise 
stated, we have performed  conjugate gradient relaxations,
keeping the pseudo--hydrogen 
atoms and the lowest two layers of C atoms fixed.
In the diffusion barrier study we have applied a constrained
conjugate gradient technique (see Fig.~\ref{fig2}).\\
We have  observed that, owing to
the relatively small size of our supercell, 
$\Gamma$ point sampling produces unphysical results.
This stems from the fact that  
at the Gamma point, the electronic states on the surface are
lower in energy compared to the bulk states,  a result which is not
generally reproduced at other k-points.
When no further k-point sampling 
is made, 
this leads, in the worst cases, to extra surface charges
of order half elementary charge/atom at some of the surface atoms.
This does not occur when an average over several representative K-points
is made.
The calculations have therefore  been performed using 
the (2 $\times$ 2 $\times$ 1) k-point-grid
recommended by Cunningham \cite{cunningham}.
\\

\iflayout
  \begin{figure} \epsfig{file=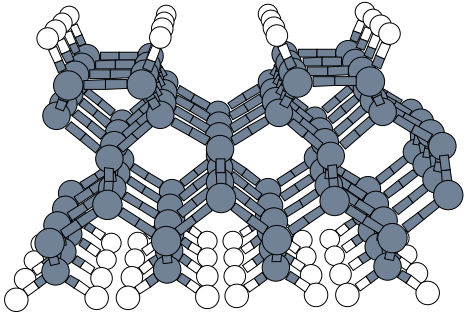,width=7.5cm} 
\caption{The model of the reconstructed diamond (100):H 2 $\times$ 1 surface.\\
}
\label{fig1}
\end{figure}
\fi  
The diffusing atom is moved stepwise from 
the starting to the final position and 
is allowed to relax in the plane perpendicular
to the direction of the vector connecting its' starting and 
 final positions. No constraints are applied 
to other atoms (except the fixed lowest two layers of C atoms).
\iflayout
  \begin{figure} \epsfig{file=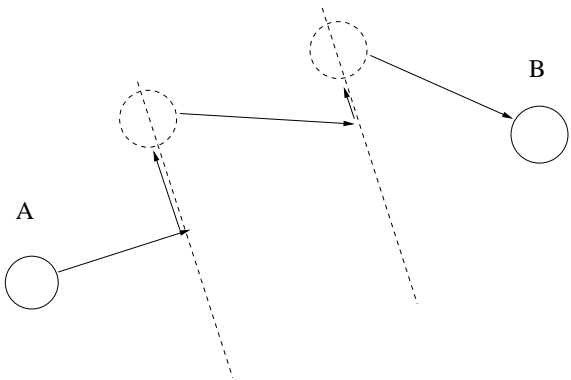,width=7.5cm} 
\caption{The constrained conjugate gradient relaxation.\\
}
\label{fig2}
\end{figure}
\fi

%\newpage
\section{The Harris Mechanism}
The initial stages of the Harris mechanism can be divided into 4 steps: 
(i) removal  of an 
H atom from an otherwise fully H-terminated surface, (ii) adsorption of
a CH$_{3}$ radical at the newly formed dangling bond site, (iii) 
loss of H from the CH$_{3}$ adsorbed species  and simultaneous formation
a C=C double bond with a surface C,  
which breaks its surface reconstruction bond  
whilst leaving the 
adjacent surface atom 3--fold coordinated.  
It can be considered that the steps (i) to (iii) inclusive are a 
complex mechanism  by which a CH$_{2}$ group is deposited in a position 
where it can ``attack'' the  weakened surface reconstruction bond. 
This is achieved  in (iv), where the 
CH$_{2}$   species rotates 
into the bridging position above the two surface C atoms.
Steps (i)-(iv) are illustrated in Fig.~\ref{fig3}.

\iflayout
  \begin{figure} \epsfig{file=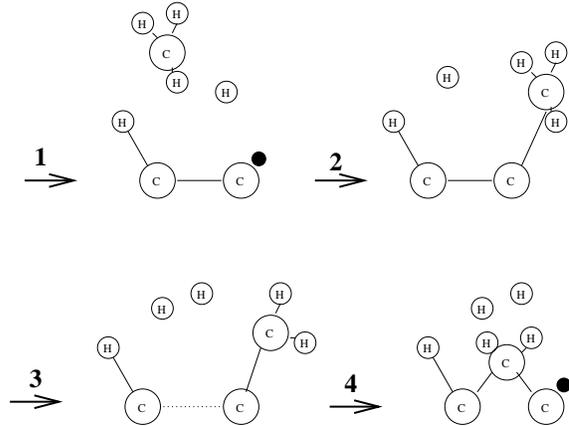,width=7.5cm} 
\caption{The initial steps in diamond growth\\
on dimerised diamond (100):H surface according to Harris.
}
\label{fig3}
\end{figure}
\fi  

We cannot accurately 
calculate barriers for processes of ad/desorption to/from
a surface, such as those in (i), (ii) and (iii), 
since charge transfer effects  within DF-TB mean that
the detaching radical--surface 
complex cannot be properly represented. However, if ad/desorption is not
accompanied by any significant electronic or structural relaxation, as indeed 
is the case in steps (i) \& (ii) for the impurity free 
surface, we can safely
assume that there are no significant additional contributions to the 
energy barriers to such processes
other than the difference in formation energy between the initial  and final
structures. As we shall describe in the next section, this is not
so for the impurity case, where structural reorganization around the N and an 
accompanying subsurface impurity--surface charge transfer  occurs.
We can therefore not
talk with any confidence about the energy barriers here. 
In the light of this,
we must limit our discussion for steps (i) to (iii), where the particle 
number at the surface is not conserved,  to comparing formation
energies for the resultant structures and making inferences where 
possible as to the nature of the energy barrier between. In the case of 
process (4), where surface particle number is conserved, 
calculation of an energy
barrier is possible within our method.

\section{Results}
We discuss here the energetics of each of the steps of the
Harris mechanism described in the previous section for 
the impurity free and the subsurface N and B calculations. 
We show  in Tab.~\ref{harris} the calculated differences
in formation energies for steps (i) to (iv) inclusive and also the energy
barrier for step (iv).
The relative energies after each step are depicted in Fig.~\ref{fig3b}.\\

\iflayout
  \begin{figure} \epsfig{file=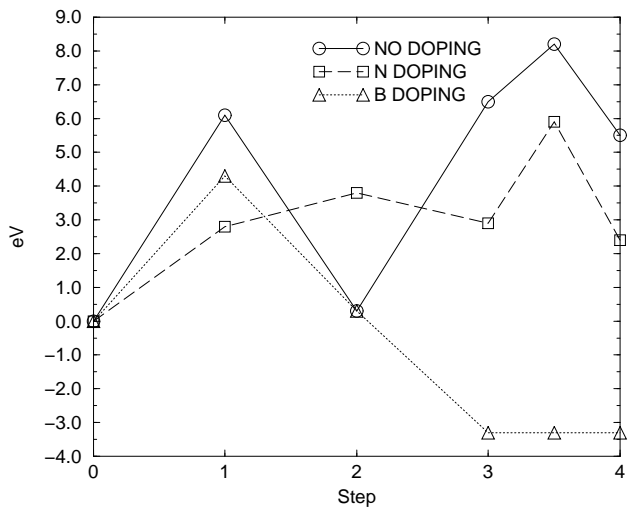,width=7.5cm} 
\caption{The relative total energies after each step (i) - (iv).\\
The zero of the energy is the energy of the
three differently doped initial structures.
The energy barrier of the step (iv) is also shown.
}
\label{fig3b}
\end{figure}
\fi 
 
${\em Step (i): Removal~of~H~from~the~surface}$.\\
We obtain 6.1 eV for the binding energy of an H atom to
the  undoped
surface. 
This high value is in agreement with other theoretical
calculations
\cite{anderson,garrison,latham} and reflects the 
strong nature of the C--H bond. The binding energy in the presence of
N,  at 2.8 eV, is much lower. This is  due to the occurrence of a 
structural relaxation after a removal of the surface H atom, 
consequently lowering the energy of the final structure:
the N atom moves  from off to onsite and an  
electron migrates from the
impurity atom to the surface. Such a  process has been described 
in detail in an 
earlier paper\cite{sitch2}, where it was shown that the position of the N 
atom in the lattice is governed by the Fermi level. Namely, 
when E$_{f}$ lies at or above the single occupied A$_{1}$ level 
associated with the defect, the N atom lowers its energy by moving offsite
along one of the bonding ${<}$111${>}$ directions. Conversely, if E$_{f}$
is pinned below A$_{1}$, onsite N is stabilized by a charge transfer to
deeper lying states. The latter is the case here: the removal of an 
H atom from the surface leaves a deep lying dangling bond state, to 
which an electron migrates from the neighbourhood of the N atom. We observed 
in\cite{sitch2} that this spontaneous onsite motion is accompanied by 
an energy gain of 1.4 eV as measured by  DF--TB.
The transference of charge to the surface  is confirmed in our 
case by a
Mulliken study, which  shows that 
a lone pair now resides on 
the 3--fold coordinated surface C atom. Thus  the 
formation energy of the resulting structure is reduced.
For B doping, the binding energy is lowered to 4.3 eV. Mulliken
studies show clearly that a similar charge transfer effect is also 
responsible here - the surface dangling bond electron is pulled into a
deep-lying subsurface acceptor state associated with the B atom.\\
${\em Step (ii): Methyl~Absorption}$.\\
The methyl radical has  the  largest binding energy, 5.84 eV,
when attaching to 
the non-doped surface,  indicating  the
strength  of the ${\sigma}$ C-C bond. 
Adsorption of the CH$_{3}$ in the presence of a subsurface N atom is not 
favored, instead of a binding energy we find that this step costs 
${\approx}$ 1 eV. 
This stems from the inherent stability of the initial structure. We also 
suggest that a large barrier will exist for this process, 
since the site to which the radical
should attach is no longer a dangling bond, as is the case for the impurity
free supercell, but a fully saturated
lone pair. 
The electrostatic repulsion between  the lone pair  and the CH$_{3}$ 
radical must first be overcome in order for a bond to be formed.
In the B doped case the CH$_3$ binding energy is lowered to 4.04 eV, 
which again can be attributed to the charge transfer induced stability 
of the start structure.\\
{\em Step 3:~H~abstraction~and~surface rearrangement}. 
The cost of extraction of an H atom from the CH$_3$ species is again 
relatively high for the impurity free case at 6.2 eV. A  C-C sp$^{2}$ bond is 
spontaneously  formed, with 
the C and H atoms in CH$_2$ and the C atom on the surface all lying
roughly
in the same plane. 
The dimer-dimer
bonding 
 close to CH$_2$ lengthens by 13\%. This weakening is crucial for the final
step in the growth process, in which the CH$_{2}$ group rotates into a 
bridging position  above this bond, breaking it in the process.
On the N doped surface, the CH$_2$ fragment maintains the
sp$^3$-like configuration, with charge transfer from the 
subsurface N to the CH$_{2}$ adspecies, thus saturating the 
newly created dangling bond in the form of  a lone pair (i.e. identical 
charge transfer mechanism to that of step (i)). 
In contrast to the undoped case, 
the surface reconstruction bond is not
lengthened. As we shall explain  in the discussion of step (iv), 
this actually
hinders growth.
For  B doping, the surface spontaneously rearranges:
the CH$_2$ group occupies the bridging position and the Harris cycle 
is completed. The energy gain in this process is 3.59 eV. \\
%As with step 1, this step is  energetically cheaper than in the
%defect free case, resulting in an energy gain of 0.88 eV.\\
{\em Step (iv):~Migration~of~CH$_{2}$~to~bridging position}.\\
We obtain an energy barrier of 
1.75 eV 
for the CH$_2$ diffusion to
the bridge position with the undoped sample, in reasonable agreement with
Anderson, who  has found 
this barrier to be less than 1.92 eV \cite{anderson}.
The N doped sample gives an energy barrier of 
3.03 eV for the CH$_2$ diffusion, which is understandable 
since in this case the
surface reconstruction bond must be broken, which is energetically costly.
For B, as previously stated, the incorporation of the 
CH$_2$ fragment to the bridging position takes place with no energy barrier. 

\section{Discussion}

\subsection{Nitrogen Doping}
It is  clear from these results that the Harris mechanism
cannot explain N catalysis of (100) diamond growth. 
 Without
doping,  the hydrogen abstraction reactions 
(i) and (iii), as well as the energy barrier for the motion of the
CH$_{2}$ adspecies  to the bridge 
position (iv), are the most prohibiting steps.
Our results suggest that step (ii),  where in the impurity free case
a CH$_{3}$ group attaches to a surface dangling bond site, is
severely hindered by the presence of subsurface N.
Here, charge transfer from N
to the 
surface means the CH$_{3}$ radical  must attack a fully saturated site, where
the C surface atom has an  
associated lone pair of electrons. The probable high energy barrier 
to overcome such an electrostatic repulsion 
suggests that the CH$_3$ bonding to the 
surface in step (ii)
is unlikely.
Further, the subsurface-surface  charge transfer 
severely disrupts step (iii). In the undoped case, the extraction
of an H atom leads to the formation of a C=C adatom-surface sp$^{2}$
bond,  
together with a weakening of the adjacent surface reconstruction bond.
In the doped case, charge transfer from the N atom to the C adatom saturates
the dangling bond, thus leaving the C-C adatom-surface bond sp$^{3}$ 
like and the
surface reconstruction bond unperturbed. 
%This then leads to a 
%high energy barrier for step (iv), since a stronger 
%reconstruction bond must be
%broken by the  adatom
% CH$_2$ group rotating into the bridge position overhead.
A critical analysis of the Harris mechanism would suggest step (iii) to be
the most crucial in the whole process, since it at once places a CH$_{2}$
group in a position where it can ``attack'' a weakened surface reconstruction
bond,  subsequently forming a bridge site, which acts as a seed for 
further growth on the plane.
This is manifestly not the case when subsurface N is present, 
where a full strength C-C reconstruction bond must be broken by 
an essentially ``saturated'' CH$_{2}$ group (the C atom having
one C-C, two C-H and an associated lone pair) rotating into the 
bridge site.   
Thus one is led to question the suitability of such a complex model in this case.
In  section  VI we describe a possible alternative.

\subsection{Boron Doping}
Although the energetics of Harris mechanism is perturbed by the 
presence of subsurface B atoms, this does not suggest that 
the mechanism should cease to be valid in this case.
Just as for N dopants,
a  charge transfer 
is responsible for the  discrepancy in the 
formation energies of the start and finish structures for steps (i) and (ii)
between the B doped and impurity free structures. However,
this does not lead to the problems encountered with  N, since 
charge is now transferred {\em from} the surface to a
subsurface B acceptor level. The structure after H abstraction
(step (i)) is stabilized by  charge transfer, with the 3-fold
coordinated surface C atom now having one completely empty level.  
Hence although adsorption of a CH$_{3}$ radical is now not as attractive 
as when a dangling bond is present (impurity free surface), there
is not, as is the case for N, 
%where a lone pair of electrons is resident at the
%target C-site, 
an  electrostatic
repulsion preventing such an occurrence. 
Once the CH$_{3}$ group is adsorbed onto the surface (step (ii)), the rest 
of the Harris mechanism is energetically favorable.
Although we  cannot say exactly how big the energy barrier for H 
abstraction from 
the CH$_{3}$  group is, we can reason that it has as its' upper bound the  
energy for abstraction 
from the undoped surface. This is  due to charge transfer  during 
abstraction - removal of H from the undoped surface requires the 
breaking of a  full
strength C-H bond, whereas when a B subsurface dopant is present, the energy
barrier for the process may be lowered by 
charge transfer to the subsurface B atom.
After H abstraction,  the 
relatively electropositive CH$_{2}$ group is pulled 
spontaneously to the electron rich 
bridge site. The overall energy gain in H abstraction + CH$_{2}$ 
diffusion to the bridge site is 3.6 eV. 

\section{An Alternative Model for Growth with N Doping: the ``Zipper'' mechanism}
%In the last subsection, it was explained how in the presence of 
%subsurface N, the rate limiting step  for growth within the Harris mechanism
%is the breaking of a full strength 
%C-C reconstruction via attack by an adsorbing species. This being the case, 
We consider a
far simpler method would be   more appropriate to describe 
N-catalysed (100) growth, 
since it would  remove the unnecessary 
and costly initial steps of the Harris mechanism.
We suggest here one such model.
We have found in our studies that, 
although the  3--fold 
coordinated N atom is the most stable configuration 
for a fully hydrogenated (100) 
surface, a structure where the excess ``doping'' charge is transferred to
a surface reconstruction ${\sigma^{\ast}}$ state is metastable. This has been
confirmed by an {\em ab-initio} 
all-electron cluster calculation, using the code developed 
by Pederson and Jackson\cite{pederson}, where  a difference in energy
of 2.40 eV between the two structures is found. The ${\sigma^{\ast}}$ state
is strongly localized on one reconstruction C-C bond, which 
as a consequence lengthens from 1.62 {\AA} to 2.30 {\AA}. We have 
found this electron rich site to be an ideal adhesion point for a 
CH$_{2}$ species. Indeed, using the {\em ab-initio} cluster code
of Jones and Briddon\cite{bob2}, we 
observe  no energy barrier for the adhesion 
process and a binding energy of ${\approx}$ 8 eV. Once the CH$_{2}$ species 
adheres to the surface, the bridging and bridged C atoms are electronically
saturated, thus allowing the ``doping'' charge to migrate to the next adhesion
site and so on. 
Growth of a whole layer may thus be catalysed by the presence of one 
N electron.

 We therefore visualize the growth process in the following way:
the growing crystal is a non-equilibrium thermodynamic system, in which 
atoms on the surface are vibrating in a variety of different phonon modes. It is 
perfectly plausible that the two carbons of a reconstruction bond describe a ``breathing mode'', 
in which their C-C bond length is periodically much larger than the already weakened C-C reconstruction 
bond. This therefore represents an ideal target for an adhering CH$_{2}$ species. The 
energy barrier to overcome the breaking of the residual C-C  reconstruction bond is further lowered by the 
simultaneous transfer of charge from the subsurface N to the surface. Once the CH$_{2}$ adhesion  at
the bridging site is completed, the excess electronic is free to mediate a similar reaction at the
adjacent site. Thus growth of a whole layer may thus be catalysed by the presence of one
N electron.\\
Due to the geometry of the diamond structure,
smooth growth in the (100) direction requires 
the dimer row on the upper terrace to be 
perpendicular to the dimer row in the lower terrace.
This can be achieved by the dimer opening reaction 
if  two CH$_{2}$ adjacent adspecies (see Fig.~\ref{fig4})
both eject one of their H atoms and bond together to form an isolated dimer.
%leaving still one dimer open in the lower terrace.
This isolated dimer can thereafter transform to a C=C$H_{2}$ adspecies 
and migrate towards an existing dimer row as proposed 
by Skokov\cite{skokov}. 
The suggested new model is depicted in Fig.~\ref{fig4}.
Instead of CH$_{2}$, the CH$_{3}$ molecule 
may also be a good candidate 
attaching to the open dimer. In this case two H$_{2}$ abstractions
are required.

\iflayout
  \begin{figure} \epsfig{file=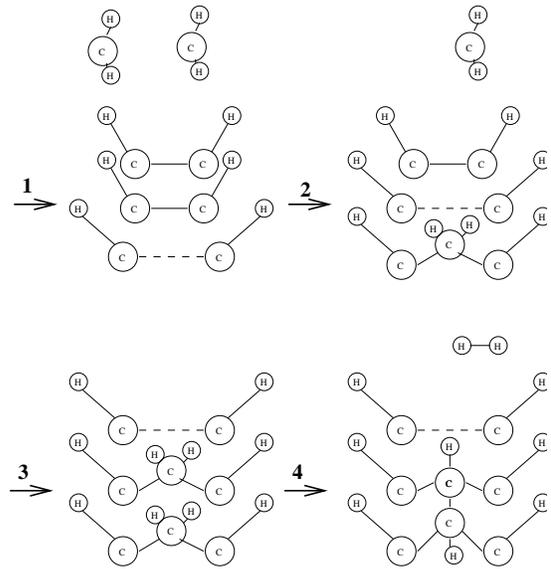,width=7.5cm} 
\caption{The novel growth model with N doping of CVD
(100):H diamond: the Zipper mechanism. 
i) The extra electron from N migrates to 
the surface and opens a dimer bond. 
ii) A CH$_{2}$ adsorbs 
to the open dimer, and the neighboring dimer is opened.
iii) Another CH$_{2}$ adsorbs 
to the open dimer, and the next dimer is opened.
iv) H$_{2}$ is abstracted and a new isolated dimer is formed
to the upper terrace.
\\
}
\label{fig4}
\end{figure}
\fi  

In our argument thus far we have neglected two important questions: (1) how
big is the energy barrier for the dimer opening? (2) why is this 
method only valid for (100) orientations?
We estimate (1) by noting that the essential 
difference between the stable 3-fold
coordinated N + closed dimer structure and that of the 
metastable 4-fold
coordinated N + open dimer consists of the energy cost of breaking the C-C 
reconstruction bond and the energy gain of the onsite motion of N on 
losing an electron. We have calculated the former to be 2.4 eV and argue 
in section IV
above that the latter 
is 1.4 eV. Hence we arrive at the energy barrier of 1.0 - 2.4 eV,
a plausible figure given the energies discussed in connection with the 
Harris mechanism.\\
Point (2) is answered by noting that the (100) differs from 
the (111) and (110) surfaces in that the clean surface possesses
 two dangling bonds per atom. Reconstruction and hydrogenation 
results in a structure where the surface C atoms have two C-C bulk,
one  C-C surface, plus a saturating C-H bond. Hydrogenated (110) 
and (111) surfaces possess 3 bulk C-C plus one C-H bond. The 
reconstruction surface (100) C-C
bond, at 1.62 {\AA} is longer,  and consequently weaker and more 
vulnerable to attack than a 
bulk ${\sigma}$ bond. In the case of, for example the 
hydrogenated (111) surface, no such reconstruction bonds exist. To activate
a surface bond would therefore require the breaking of a far stronger 
bulk-like ${\sigma}$ bond, which is then correspondingly 
energetically more expensive and hence less probable.

\section{Conclusions}
In this paper we  have employed a density functional 
method to investigate the effect of N and B doping
on the growth of CVD diamond (100):H 2 $\times$ 1 surfaces. 
Consistent with recent CVD experiments which have shown that Boron improves the
xcrystalline quality of (100) CVD diamond surfaces, 
we have found the Harris mechanism to be an energetically 
favorable pathway in the CVD growth of 
B doped samples. 
In the N doping case, we argue that the increased 
diamond growth rate in the (100) direction cannot be accounted for
by the Harris mechanism, rather we suggest an alternative model in which 
the (100) surface is charged by N-donor electrons.
In this model CH$_{2}$ group is directly inserted into the 
bridging position.

%\vspace{-5.0cm}

\subsection*{Table caption}
\noindent

\refstepcounter{table} \label{harris}
{\bf Table~\ref{harris}:}
Differences in formation energies for the various steps in the Harris procedure, plus the energy barrier for step (iv) for (i) impurity
free (ii) subsurface N (100) surface.\\
\begin{table}
  \begin{tabular}{|l |l |l |l |l |||l |}\hline
\cline{1-2}
Step           & 1     & 2     & 3     & 4     & Energy Barrier\\ 
  &     &      &      &      & in 4 \\ \hline \hline
No doping       & +6.1  & -5.8  & +6.2  & -1.0  & 1.7   \\ \hline
N doping        & +2.8  & +1.0  & -0.9  & -0.5  & 3.0   \\ \hline
B doping        & +4.3  & -4.0  & 
\multicolumn{2}{c|||}{-3.6} & 0.0 \\ \hline
  \end{tabular}
\end{table}

\newpage
\end{document}